\documentclass[namedreferences,natbib, linksfromyear]{solarphysics}
\usepackage[optionalrh,solaromanenum]{spr-sola-addons} 
\usepackage{graphicx}                    
\usepackage{amssymb}                    
\usepackage{color}                       
\usepackage{url}                         
\usepackage[pdfborder={0 0 0},colorlinks=true,linkcolor=black,citecolor=black,filecolor=black,urlcolor=blue,breaklinks=true,dvipdfm]{hyperref} 

\usepackage{solaheader}	
\newcommand{\solareceiveddate}{1December 2014} 
\newcommand{\solaaccepteddate}{12 January 2015}
\begin{document}

\begin{article}

\begin{opening}

\title{On the Effect of the Interplanetary Medium on Nanodust Observations by the \textit{Solar Terrestrial Relations Observatory}.}

%
\author{G.~\surname{Le Chat}$^{1,2,3}$\sep
        K.~\surname{Issautier}$^{1}$\sep
        A.~\surname{Zaslavsky}$^{1}$\sep
        F.~\surname{Pantellini}$^{1}$\sep
        N.~\surname{Meyer-Vernet}$^{1}$\sep    
        S.~\surname{Belheouane}$^{1}$\sep
        M.~\surname{Maksimovic}$^{1}$\
       }

%
\runningauthor{G. Le Chat {\it et al.}}
\runningtitle{Effect of the Interplanetary Medium on Nanodust Observations.}

%
  \institute{$^{1}$ LESIA-Observatoire de Paris, CNRS, UPMC Univ. Paris 6, Univ. Paris Diderot; 5 Place Jules Janssen, 92195 Meudon, France\\ 
                     email: \url{gaetan.lechat@obspm.fr}\\ 
             $^{2}$ NASA Lunar Science Institute, Moffet Field, CA, USA\\
             $^{3}$ Harvard-Smithsonian Center for Astrophysics, Cambridge, USA\\
}

\begin{abstract}
Dust particles provide an important fraction of the matter composing the interplanetary medium, their mass flux at 1 AU being comparable to the one of the solar wind. Among them, dust grains of nanometer size-scale can be detected using radio and plasma wave instruments, because they move at roughly the solar wind speed. The high velocity impact of a dust particle generates a small crater on the spacecraft: the dust particle and the crater material are vaporized. This produces a plasma cloud whose associated electrical charge induces an electric pulse measured with radio and plasma instruments. Since their first detection in the interplanetary medium, nanodust particles have been routinely measured using the {\it Solar Terrestrial Relations Observatory} /WAVES experiment [S/WAVES]. We present the nanodust properties measured using S/WAVES/\textit{Low Frequency Receiver} [LFR] observations between 2007 and 2013, and for the first time, present evidence of coronal mass ejection interaction with the nanodust, leading to a higher nanodust flux measured at 1 AU. Finally, possible influences of the inner planets on the nanodust flux are presented and discussed.

\end{abstract}

%
\keywords{Interplanetary Dust, Nanodust, In situ Dust Detection, Radio Antennas, STEREO/WAVES, Solar Wind, Coronal Mass Ejections}

\end{opening}

%
 \section{Introduction}

Interplanetary nanodust have been detected at 1 astronomical unit [AU] \citep{Meyer09, Zaslavsky12, LeChat13} by a radio- and plasma-wave instrument, the {\it Solar Terrestrial Relations Observatory} /WAVES experiment [S/WAVES] \citep{Bougeret08}. A confirmation of this detection has been recently made using the \textit{Cassini/Radio and Plasma Wave Science} [RPWS] instrument \citep{Schippers14}. This detection is made possible by the interaction between the nanodust and the interplanetary plasma and magnetic field, which accelerate them up to the solar wind speed due to their relatively large charge-over-mass ratio \citep{Mann2010}. The amplitude of the signal induced by a dust grain impacting a spacecraft grows much faster with speed than with the mass of the dust particle \citep{McBride99}. As a result, nanodust can produce a strong signal, despite their low mass \citep{Zaslavsky12, Pantellini13}.

Interactions between magnetized plasma and nanodust determine the dynamics of the nanoparticles, which made the detection of the nanodust possible and allow these particles to escape the Sun's gravity \citep{Czechowski10, Mann2010, Juhasz13, Belheouane14}. Concepts of probing the interplanetary magnetic field using nanodust stream particles in Saturn's magnetosphere have also been proposed \citep{Hsu11, Hsu13}.

In the present paper, after describing nanodust observations and providing a statistical analysis of the nanodust flux measured by S/WAVES \textit{Low Frequency Receiver} [LFR] between 2007 and 2013 (see Section \ref{s:obs}), we study the effects on nanodust fluxes of transient events in the solar wind, such as coronal mass ejections and stream interaction regions, and show for the first time that coronal mass ejections can interact with interplanetary nanodust leading to a higher nanodust flux measured at 1 AU (Section \ref{s:cme}). Finally, evidence of effects of inner planets on the observed nanodust flux are shown and discussed (Section \ref{s:period}).
     
 \section{Nanodust Observations}\label{s:obs}
 
 	Despite the radio experiment on STEREO was not designed to measure dust, nanodust particles are observed by two of the S/WAVES instruments: (i) the {\it Low Frequency Receiver} [LFR], which is a spectrum analysis receiver observing from 2.5 kHz to 160 kHz designed to measure the ubiquitous plasma thermal noise and solar radio bursts; ii) the {\it Time Domain Sampler} [TDS] which makes high-rate samples of waveforms primarily to study Langmuir waves \citep{Bougeret08}. Detailed analyses of nanodust measurements were made by \citet{LeChat13} using LFR, and by \citet{Zaslavsky12} using TDS. 
     
     When nanodust is observed on STEREO-A, the impact zone is close to the X-antenna, producing a voltage pulse typically 20 times higher than in the two other antenna booms as shown by the TDS measurements \citep{Zaslavsky12}. The voltage pulse on the X-antenna is given by \citet{Pantellini13}
     \begin{equation}\label{va}
     	\delta V_\mathrm{X}\,=\,\Gamma \frac{Q_\mathrm{a}(l)}{C_\mathrm{a}},
     \end{equation}
 \noindent with 
 \begin{equation}\label{Qa}
 	Q_\mathrm{a} (l) \approx j_\mathrm{ph, 1 AU} 2r_0 l \tau ,
 \end{equation}
 \noindent  the charge of photoelectrons emitted during a time $\tau$ by a boom of length $l$ and radius $r_0$, with $j_\mathrm{ph, 1 AU}$ the photoelectron current density at 1 AU. $\Gamma\,\approx\,0.5$ and $C_\mathrm{a}\,=\,60 \mathrm{pF}$ are the antenna's gain and capacitance \citep{Bale08}. Assuming equipartition between the average potential energy of the photoelectrons bound to a boom of length $l$ (and total charge $Q_\mathrm{a} (l)$) and their thermal energy, Equation (\ref{va}) can be simplified to
      \begin{equation}\label{VX}
     	\delta V_\mathrm{X}\,\approx\, \frac{\Gamma T l}{L},
     \end{equation}

 \noindent where $L\,=\, 6\,\mathrm{m}$ is the antenna's length \citep{Bale08}, $l$ is the length of the antenna within the plasma cloud, and $T\,=\,2.5\,\mathrm{eV}$ is an effective temperature given in eV, of the order of the photoelectron temperature \citep{Zaslavsky12, Pantellini12,Pantellini13, LeChat13}. The parameter $l$ in Equation (\ref{Qa}) and (\ref{VX}) implies that there is a critical radius $R_\mathrm{C}$ up to which the antenna is affected by the plasma cloud created by the impact. The existence of $R_C$ has been extensively demonstrated in \citet{Zaslavsky12}. The critical radius $R_\mathrm{C}$ can be interpreted as the radius for which the electron number density of the ambient plasma is large enough to screen the charge created by the impact, whose electric field perturbs the antenna's photoelectrons \citep{Pantellini13}. $R_\mathrm{C}$ is given by $R_\mathrm{C}\,\approx\,(3Q/4\pi e n_\mathrm{a})^{1/3}$, where $Q$ is the charge of the cloud created by the dust impact on the spacecraft, and $n_\mathrm{a}$ is the ambient electron density. $Q$ is given with a large uncertainty by \citep{McBride99}
 \begin{equation}\label{Q}
 	Q\ \approx\ 0.7 m v^{3.5}.
 \end{equation}
 
\citet{LeChat13} have shown that the rise time [$\tau$] and the amplitude [$\langle N\delta V^2 \rangle$] (where $N$ is the number of impacts per time unit) obtained from the fitting of the LFR spectra of STEREO-A allow to infer the nanodust flux observed at 1 AU, from the following relation
 
 \begin{equation}\label{fluxNVT}
	\langle N \delta V^2/\tau^2\rangle\ \approx\ f_0 \pi K^4 m_\mathrm{R_{SC}}^{-\gamma+7/3} \left( \frac{\Gamma T}{L\tau}\right)^2,
\end{equation}

\noindent where $K\,=\,R_\mathrm{C}/m^{1/3}\,\approx\,(3\times0.7 v^{3.5}/4\pi n_\mathrm{a} e) ^ {1/3} $.  $m_\mathrm{R_{SC}}\,=\,(R_\mathrm{SC}/K)^3$ is the mass of the impacting dust for which $R_\mathrm{C}\,=\,R_\mathrm{SC}\,=\,0.84\,\mathrm{m}$, with $R_\mathrm{SC}$ the effective size of the spacecraft \citep{Zaslavsky12, LeChat13}. The flux of particles of mass between $m$ and $m\,+\,\mathrm{d}m$ is given by $f(m)\mathrm{d}m\,=\,f_0m^{-\gamma}\mathrm{d}m$, where $\gamma$ is assumed to be equal to 11/6 corresponding to collisional equilibrium. While assuming a value for $\gamma$ is necessary to obtain numerical values of the nanodust flux measured by STEREO-A/WAVES LFR, deviation from collisional equilibrium can be expected. 

In Equation (\ref{fluxNVT}), both $K$ and $m_\mathrm{R_{SC}}$ are functions of the ambient electron density [$n_\mathrm{a}$]. This leads to the following relation between LFR dust signals [$\langle N \delta V^2/\tau^2\rangle$] and $n_\mathrm{a}$ :

\begin{eqnarray}\label{NVTDens}
	\langle N \delta V^2/\tau^2\rangle & \approx & f_0 \pi k^{3(\gamma -1)} R_\mathrm{SC}^{7-3\gamma} \left( \frac{\Gamma T}{L\tau}\right)^2 n_\mathrm{a}^{1-\gamma}\nonumber \\
	& \approx & f_0 \pi k^{5/2} R_\mathrm{SC}^{3/2} \left( \frac{\Gamma T}{L\tau}\right)^2 n_\mathrm{a}^{-5/6},\ \mathrm{for}\ \gamma\,=\,11/6,
\end{eqnarray}

\noindent with $k\,\approx\,(3\times0.7 v^{3.5}/4\pi e)^{1/3}\, \approx\, 8\times 10^8$ for a typical value of the impact speed $v\,=\,300\,\mathrm{km\,s^{-1}}$. Two electron populations contribute to the ambient density [$n_\mathrm{a}$]: the spacecraft photoelectrons and the solar wind plasma.  Due to the anti-sunward position of the antenna on the STEREO spacecraft \citep{Bougeret08, Bale08}, to the expansion of the cloud away from the spacecraft surface, and the absence of direct measurements, we cannot estimate precisely $n_\mathrm{a}$.

  \begin{figure} 
 \centerline{\includegraphics[width=0.5\textwidth,clip=]{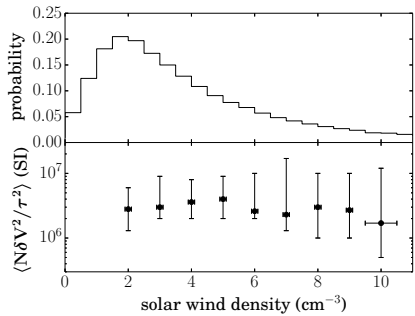}}
 \caption{Histogram of the solar wind density measured by STEREO-A/PLASTIC (top), and mode and full width at half maximum of dust signals observed by STEREO-A/WAVES LFR \textit{versus} the solar wind density (bottom). }\label{fig:corr_all}
 \end{figure}

Figure \ref{fig:corr_all} (bottom) shows the value of $\langle N \delta V^2/\tau^2\rangle$ observed by STEREO-A/WAVES LFR from January 2007 to December 2013 as a function of the solar wind  density measured by PLASTIC. The values plotted are the most probable value [mode] and the full width at half maximum [FWHM] observed for a given solar wind density. The large FWHM is primarily due to the variable nature of  the nanodust flux seen by STEREO-A in the interplanetary medium \citep{Zaslavsky12, LeChat13}, with variations spanning three orders of magnitude, with a very asymmetric distribution. Figure \ref{fig:corr_all} (top) also provides the histogram of the solar wind density to illustrate the increasing statistical error of the mode of $\langle N \delta V^2/\tau^2\rangle$ as density increases due to the lack of measurements at high solar wind density. As one can see from Figure \ref{fig:corr_all} the observed signal does not vary as expected with the solar wind density. Neither the value of $n_\mathrm{a}$ nor its variations could be precisely estimated, and in this paper, as in the previous study, we take $n_\mathrm{a}\ =\ 5\ \mathrm{cm}^{-3}$ but assume that it may go up to ten times more while estimating the error of the measurements given in Table \ref{tbl:1}.

 \begin{table}
 \caption{Flux of nanodust observed at 1 AU by STEREO-A/WAVES LFR. The systematic errors correspond to the worst cases (see text). }\label{tbl:1}
 \begin{tabular}{l c c}     
 \hline
  & $f_0$ & $F_\mathrm{10^{-20}\,kg}$ \\
  & $[\mathrm{kg^{-1}\,m^{-2}\, s^{-1}}]$ & $[\mathrm{m^{-2}\, s^{-1}}]$\\
 \hline
  Mean (lower limit) & $2.9\times 10^{-18}$ & 0.16\\
  Mean (upper limit) & $2.4\times 10^{-17}$ &1.34\\
  Standard deviation &  $3.6\times 10^{-17}$ & 2\\
  Mode & $2.5\times 10^{-19}$ &  $1.4\times 10^{-2}$\\
  FWHM & $[1.42\times 10^{-19} ; 1.1\times 10^{-18}]$ & $[8\times 10^{-3} ; 5 \times 10^{-2}]$\\
 Systematic error (underestimation) &  $1.1\times 10^{-16}$ & 6.2\\ 
 Systematic error (overestimation) &  $-2\times 10^{-19}$ & $-1.1\times 10^{-2}$\\ 
   \hline
 \end{tabular}
 \end{table}
 
Table \ref{tbl:1} provides statistical values (means, standard deviation, mode and full width at half maximum [FWHM]) of the flux of nanodust obtained using STEREO-A/WAVES LFR measurements. However, due to the sporadic behavior of the STEREO-A/WAVES nanodust measurements \citep{Zaslavsky12, LeChat13}, it is only possible to calculate two extreme values for the mean. The upper value is obtained by considering only the $651\,246$ spectra with dust measurements. The lower value is obtained by assuming that there is no nanodust when not measured by STEREO-A, \textit{i.e.} assuming $f_0\ =\ 0$ for the $4\,779\,857$ spectra without dust signals observed between January 2007 and December 2013. The upper value must be an overestimate since, as stated by \citet{LeChat13}, LFR cannot observe individual nanodust impacts or values of $\langle N \delta V^2/\tau^2\rangle$ smaller than the quasi-thermal noise of the solar wind plasma. On the other hand, the lower value may be an underestimate due to the effect of the spacecraft geometry and of the dynamic properties of the nanodust which in some interplanetary field configurations defocused the nanodust away from the ecliptic plane where STEREO-A is orbiting \citep{Belheouane14, Juhasz13}. 

In Equation (\ref{NVTDens}), the parameter $k$, the effective temperature $T$ and the density $n_\mathrm{a}$ are not measured but estimated to a constant value. This leads to systematic errors on the flux measurements, for which two worst cases can be calculated and are given in Table \ref{tbl:1}. These errors can be directly added to the values of the upper limit mean, mode and FWHM, but must be multiplied by the ratio of dust measurements over the number of spectra (\textit{i.e.} 0.12) before being added to the lower limit mean. The largest underestimation of the nanodust fluxes is obtained using the smallest possible value of $k$ and $T$, and the largest density $n_\mathrm{a}$. The largest overestimation is obtained with the largest value of $k$ and $T$, and the lowest possible ambient density. These errors are dominated by the uncertainty of the value of the parameter $k$. The error on the parameter $k$ can be estimated to be of an order of magnitude, and comes from both the uncertainty on the charge $Q$ created by the impacting dust (Equation (\ref{Q})) and of the relative speed of the dust relative to the spacecraft due to changes in the direction of the dust speed. It is noteworthy that this latter uncertainty plays also a significant role in the observed standard deviation. 

\section{Effect of Transient Events on Dust Measurements}\label{s:cme}

Transient events in the solar wind, like interplanetary coronal mass ejections [ICME] and stream interaction regions [SIR], change locally the solar wind and interplanetary magnetic field properties. These modifications would change the dynamic behavior of already released nanodust, or might allow part of the trapped nanodust below 0.2 AU from the sun to be accelerated outwards \citep{Czechowski10, Mann2010}.  We have determined the transient events observed by STEREO-A using the detection mechanism for ICMEs and SIRs given by \citet{Jian06a, Jian06b, Jian13}.  

  \begin{figure} 
 \centerline{\includegraphics[width=0.5\textwidth,clip=]{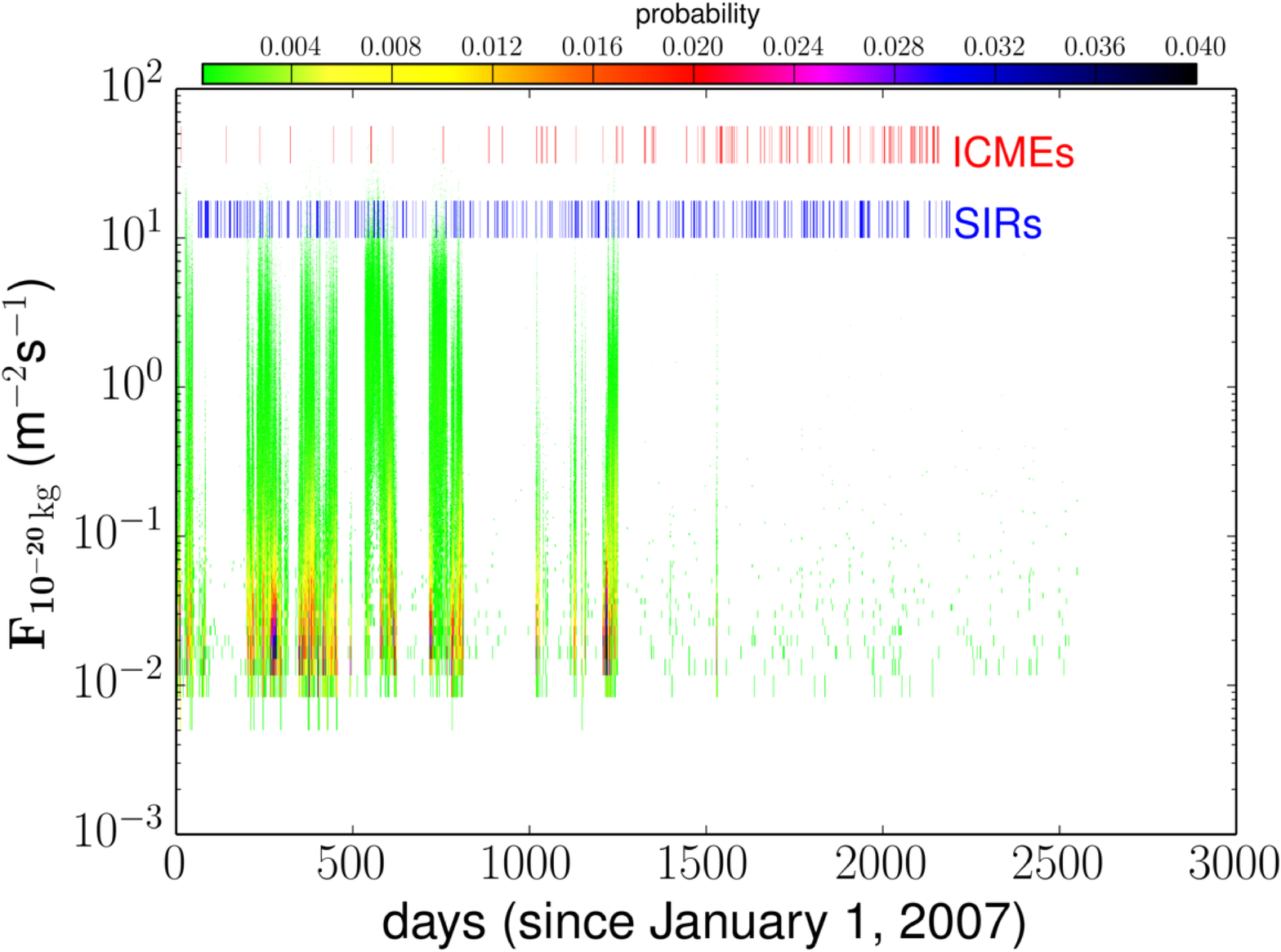}}
 \caption{Variation with time of the cumulative flux of particles of mass greater than $10^{-20}$ kg measured by STEREO-A/WAVES LFR between 2007 and 2013. Red vertical bars correspond to the time periods when STEREO-A encounters ICMEs, while blue ones indicate SIR events.}\label{fig:flux_time}
 \end{figure}
 
Figure \ref{fig:flux_time} compares periods when STEREO-A was within transients events to the cumulative flux of particles heavier than $10^{-20}$ kg: red vertical bars correspond to the time periods when STEREO-A encounters ICMEs, while blue ones indicate SIR events. No time correlation between either ICMEs or SIR events and the observed flux of nanodust is observed. During solar activity maximum, when the number of observed ICMEs increased in STEREO-A data, the number of spectra with nanodust signatures decreased significantly, but studies of the dynamics of the nanodust suggest that this is due to the defocusing configuration of the interplanetary magnetic field, pushing away from the ecliptic the nanodust released in the inner heliosphere \citep{Juhasz13}.
 
  \begin{figure} 
 \centerline{\includegraphics[width=0.5\textwidth,clip=]{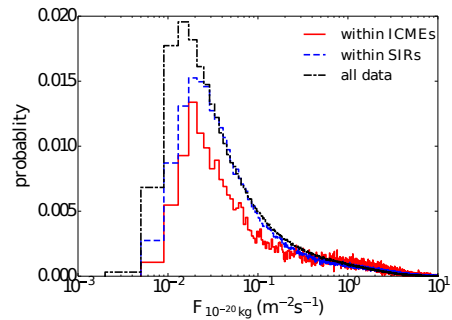}}
 \caption{Histogram of the cumulative flux of particles heavier than $10^{-20}$ kg for all dust measurements (black dash-dotted line), compared to measurements within ICMEs (red solid line) and within SIRs (blue dashed line). Note the log-scale on the X-axis.}\label{fig:hist_cme}
 \end{figure}
 
Figure \ref{fig:hist_cme} shows histograms of $F_{10^{-20}\mathrm{kg}}$ measured by STEREO-A within ICMEs (red solid line), within SIRs (blue dashed line), and for all dust measurements (black dash-dotted line).  

The distribution of the measurements within SIRs has a mode at $F_\mathrm{10^{-20}\,kg}\ =\ 1.8\times 10^{-2}\,\mathrm{m^{-2}\,s^{-1}}$, compared to the one of the whole dataset at $F_\mathrm{10^{-20}\,kg}\ =\ 1.3\times 10^{-2}\,\mathrm{m^{-2}\,s^{-1}}$. The nanodust flux within SIRs is slighty higher than for the whole dataset; however the distribution of the flux higher than $3\times 10^{-2}\,\mathrm{m^{-2}\,s^{-1}}$ is similar to the one obtained every S/WAVES LFR measurements.

\begin{figure}
 \centerline{\includegraphics[width=0.5\textwidth,clip=]{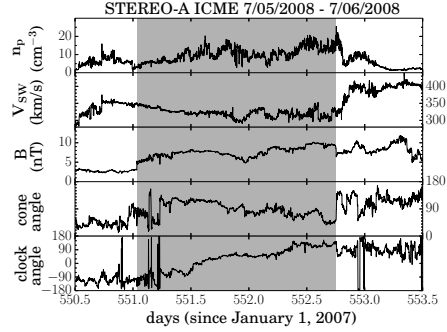}}
 \caption{Density, bulk speed, magnetic field strength, magnetic cone angle and magnetic clock angle measured by STEREO-A during the ICME between July 5, 2008 and July 6, 2008 (gray area).}\label{fig:cme6}
 \end{figure}

\begin{figure} 
 \centerline{\includegraphics[width=0.5\textwidth,clip=]{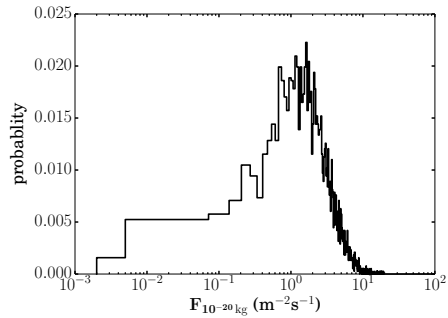}}
 \caption{Histogram of the cumulative flux of particles heavier than $10^{-20}$ kg within the ICME detected by STEREO-A between July 5, 2008 and July 6, 2008. Note the log-scale on the X-axis.}\label{fig:hist_cme6}
 \end{figure}

A bimodal distribution appears for nanodust measurements within ICMEs, the first mode at $F_\mathrm{10^{-20}\,kg}\ =\ 1.8\times 10^{-2}\,\mathrm{m^{-2}\,s^{-1}}$ and the second at $F_\mathrm{10^{-20}\,kg}\ =\ 2\,\mathrm{m^{-2}\,s^{-1}}$. The values of both modes of the cumulative fluxes are larger than the one observed for the whole dataset. Futhermore, the proportion of measurement corresponding to fluxes larger than $1\ \mathrm{m^{-2}\,s^{-1}}$ is more important for the measurements within ICMEs than for either the whole dataset or the measurements within SIRs. The second peak of the distribution comes from four ICMEs with plasma and magnetic field properties similar to the ones represented in Figure \ref{fig:cme6} for the ICME detected by STEREO-A between July 5, 2008 and July 6, 2008.  The distribution of the nanodust cumulative flux within these four ICMEs has similar properties than the one plotted in Figure \ref{fig:hist_cme6}, corresponding to the July 5-6, 2008 ICME. These four ICMEs have the following common plasma properties: i) a bulk speed between 300 and 400 $\mathrm{km\,s}^{-1}$; ii) a magnetic clock angle in RTN coordinates going from $-90^\circ$ or slightly lower (\textit{i.e.} mainly aligned with the normal direction pointing southward) to $90^\circ$ or slightly higher (\textit{i.e.} mainly aligned with the normal direction pointing northward), as seen in the bottom panel of Figure \ref{fig:cme6}. No influence of the magnetic filed strength within the ICMEs have been found. The six other ICMEs, observed by STEREO-A while measuring nanodust, show distribution of the cumulative flux with only one mode around $F_\mathrm{10^{-20}\,kg}\ =\ 1.8\times 10^{-2}\, \mathrm{m^{-2}\,s^{-1}}$, and have typical speed higher than $400$ $\mathrm{km\,s}^{-1}$, and do not show the magnetic field rotation within the T-N plane from $-90^\circ$ to $90^\circ$ observed in the four others. 

While the physical process at the origin of the higher nanodust fluxes observed within these ten ICMEs is not yet understood, our hypothesis is the following: first, nanodust are accelerated by a focusing interplanetary magnetic field [IMF] with a speed close to the ICMEs' ones, allowing the dust to interact with the plasma and magnetic field of the ICMEs. For the ICMEs whose inner magnetic field is in a focusing configuration, the interacting nanodust are trapped within the ICME, leading to the observed higher nanodust fluxes. This process also explains the absence of nanodust observed within ICMEs outside focusing  IMF configuration. This is for instance illustrated in Figure \ref{fig:flux_time} between 2011 and 2013 where the number of ICMEs increased while nanodust observations nearly disappear due to the large scale configuration of the IMF.

 \section{Effect of the Inner Planets on the Dust Flux Measured by STEREO-A}\label{s:period}
 
\begin{figure} 
 \centerline{\includegraphics[width=0.5\textwidth,clip=]{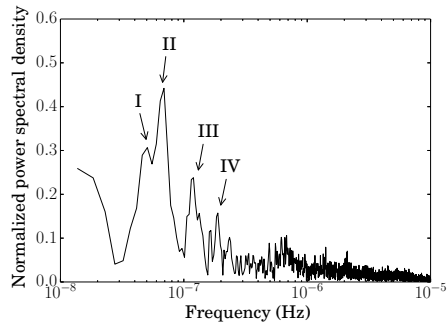}}
 \caption{Periodogram of the cumulative flux of particles heavier than $10^{-20}$ kg measured by STEREO-A/WAVES LFR. The labeled peaks of the normalized spectral power density are discussed in section \ref{s:period}. Note the log-scale on the X-axis.}\label{fig:period}
 \end{figure}

 As one can see on Figure \ref{fig:flux_time} and in \citet{Zaslavsky12} and \citet{LeChat13}, the nanodust flux observed by STEREO-A/WAVES LFR is highly variable and some periodic patterns can be seen. Figure \ref{fig:period} is the periodogram of the cumulative flux of particles heavier than $10^{-20}$ kg. Due to the uneven temporal sampling of S/WAVES LFR, the periodogram has been computed using a Lomb-Scargle periodogram algorithm \citep{Lomb76, Scargle82, Townsend10}. Four significant frequencies appear in the periodogram, labelled from I to IV from lowest to highest frequency in Figure \ref{fig:period}.
 
 The most significant peak, labelled II in Figure \ref{fig:period}, appears at a frequency of $(6.9\pm 0.3)\times 10^{-8}$ Hz. This frequency corresponds to the inverse of  half the orbital period of STEREO-A. It can be explained by the dynamic behavior of the nanodust and the large-scale IMF focusing configuration between 2007 and 2010; During this time the spacecraft has crossed the large-scale neutral line every half-rotation, as expected based on simulations of the dynamics of the interplanetary nanodust in the inner solar system \citep{Juhasz13, Belheouane14, Mann2014}. 
 
 Three other peaks, labelled I, III and IV, appear at the following frequencies: $(5.1\pm 0.3)\times 10^{-8}$ Hz, $(1.20\pm 0.03) \times 10^{-7}$ Hz, and $(1.90\pm 0.03) \times 10^{-7}$ Hz, respectively. Peak I is close to the inverse of Venus' orbit period ($5.15\times 10^{-8}$ Hz), peak III is close to the inverse of Mercury's orbit period ($1.32\times 10^{-7}$ Hz), and peak IV is close to the inverse of Mercury's rotation period ($1.97\times 10^{-7}$ Hz). While it is possible that these are coincidental, they may also represent clues of either interaction between the inner planets and the nanodust (or the dust particles responsible for the observed nanodust) or that these planetary bodies are sources of interplanetary nanodust. It is noteworthy that these frequencies are related to the planets' orbit period and not the synodic periods from the point of view of STEREO-A (which frequencies would be $1.8\times 10^{-8}$ Hz, and $9.8\times 10^{-8}$ Hz for Venus and Mercury, respectively). It implies that STEREO-A does not observe nanodust streams coming from these two planets, unlike the ones observed near outer planets (see \citet{Hsu12} and references therein), but that both Venus and Mercury increased the number of nanodust in the interplanetary medium around a fixed point on their respective orbits. A simple explanation could be the perturbation of cometary trails crossing these planets' orbits which leads to a higher collision rate locally producing more nanoparticles. The last peak corresponds to the Mercury rotation period and may be due to a hot spot on the surface of Mercury releasing dust into the interplanetary medium when illuminated by the Sun.
 
 \section{Conclusion}
 
 A major result of the work presented here is the observation of the effect of ICMEs on the flux of nanodust measured at 1 AU (see Section \ref{s:cme}). This effect is another example of dynamical influence of the magnetized solar wind on nanodust particles, which includes the acceleration of the nanodust to speed close to the solar wind speed \citep{Czechowski10, Mann2010, Juhasz13, Belheouane14}, and the increase of the nanodust stream flux ejected by the Saturn's magnetosphere \citep{Hsu11, Hsu12, Hsu13}. Our analysis shows that ICMEs can trap nanodust that has been already accelerated, but do not allow to measure nanodust at 1 AU with STEREO-A when the large-scale interplanetary magnetic field is in the defocusing configuration \citep{Juhasz13}. Future nanodust dynamic simulations coupled with semi-empirical full-MHD simulations of the solar corona and solar wind will provide significant insight into the interaction between the nanoparticles and the interplanetary medium, and will allow investigation of the effect of the ICMEs observed in our study, and testing the process suggested in Section \ref{s:cme}.
 
 In addition to the interaction with the plasma and magnetic field of the interplanetary medium, the nanodust flux observed by STEREO-A at 1 AU may be influenced by Venus and Mercury (see Section \ref{s:period}). Our result indicates that both planets increase the number of nanoparticles in the interplanetary medium as the planetary bodies encounter regions of higher interplanetary dust density, such as cometary trails.

%

%

%

%
 \begin{acks}
We thank the team who designed and built the S/WAVES instrument. The S/WAVES data used here are produced by an international consortium of the Observatoire de Paris (France), the University of Minnesota (USA), the University of California Berkeley (USA), and NASA Goddard Space Flight Center (USA). The French contribution is funded by CNES and CNRS, and the USA institutions are funded by NASA. We thank the STEREO PLASTIC Investigation (A.B. Galvin, PI) and NASA Contract NAS5-00132 for providing the proton data, and the STEREO IMPACT (J. Luhman, PI) for providing the magnetic field data. 
 \end{acks}

%
%
 \bibliographystyle{spr-mp-sola}
 \bibliography{nanoDustPlasma}  

\begin{thebibliography}{23}
\ifx \bisbn   \undefined \def \bisbn  #1{ISBN #1}\fi
\ifx \binits  \undefined \def \binits#1{#1}\fi
\ifx \bauthor  \undefined \def \bauthor#1{#1}\fi
\ifx \batitle  \undefined \def \batitle#1{#1}\fi
\ifx \bjtitle  \undefined \def \bjtitle#1{\textit{#1}}\fi
\ifx \bvolume  \undefined \def \bvolume#1{\textbf{#1}}\fi
\ifx \byear  \undefined \def \byear#1{#1}\fi
\ifx \bissue  \undefined \def \bissue#1{#1}\fi
\ifx \bfpage  \undefined \def \bfpage#1{#1}\fi
\ifx \blpage  \undefined \def \blpage #1{#1}\fi
\ifx \burl  \undefined \def \burl#1{\textsf{#1}}\fi
\ifx \href  \undefined \def \href#1#2{\textsf{#2}}\fi
\ifx \doiurl  \undefined \def
  \doiurl#1{\href{http://dx.doi.org/#1}{\textsf{#1}}}\fi
\ifx \betal  \undefined \def \betal{\textit{et al.}}\fi
\ifx \binstitute  \undefined \def \binstitute#1{#1}\fi
\ifx \bctitle  \undefined \def \bctitle#1{#1}\fi
\ifx \beditor  \undefined \def \beditor#1{#1}\fi
\ifx \bpublisher  \undefined \def \bpublisher#1{#1}\fi
\ifx \bbtitle  \undefined \def \bbtitle#1{\textit{#1}}\fi
\ifx \bedition  \undefined \def \bedition#1{#1}\fi
\ifx \bseriesno  \undefined \def \bseriesno#1{\textbf{#1}}\fi
\ifx \blocation  \undefined \def \blocation#1{#1}\fi
\ifx \bsertitle  \undefined \def \bsertitle#1{\textit{#1}}\fi
\ifx \bsnm \undefined \def \bsnm#1{#1}\fi
\ifx \bsuffix \undefined \def \bsuffix#1{#1}\fi
\ifx \bparticle \undefined \def \bparticle#1{#1}\fi
\ifx \barticle \undefined \def \barticle#1{}\fi
\ifx \botherref \undefined \def \botherref#1{}\fi
\ifx \url \undefined \def \url#1{\textsf{#1}}\fi
\ifx \bchapter \undefined \def \bchapter#1{}\fi
\ifx \bbook \undefined \def \bbook#1{}\fi
\ifx \bcomment \undefined \def \bcomment#1{#1}\fi
\ifx \oauthor \undefined \def \oauthor#1{#1}\fi
\ifx \citeauthoryear \undefined \def \citeauthoryear#1{#1}\fi
\def \endbibitem {}
\ifx \bconflocation  \undefined \def \bconflocation#1{#1} \fi

\bibitem[\protect\citeauthoryear{{Bale} \textit{et~al.}}{2008}]{Bale08}
\begin{barticle}
\bauthor{\bsnm{{Bale}}, \binits{S.D.}},
\bauthor{\bsnm{{Ullrich}}, \binits{R.}},
\bauthor{\bsnm{{Goetz}}, \binits{K.}},
\bauthor{\bsnm{{Alster}}, \binits{N.}},
\bauthor{\bsnm{{Cecconi}}, \binits{B.}},
\bauthor{\bsnm{{Dekkali}}, \binits{M.}},
\bauthor{\bsnm{{Lingner}}, \binits{N.R.}},
\bauthor{\bsnm{{Macher}}, \binits{W.}},
\bauthor{\bsnm{{Manning}}, \binits{R.E.}},
\bauthor{\bsnm{{McCauley}}, \binits{J.}},
\bauthor{\bsnm{{Monson}}, \binits{S.J.}},
\bauthor{\bsnm{{Oswald}}, \binits{T.H.}},
\bauthor{\bsnm{{Pulupa}}, \binits{M.}}:
\byear{2008},
\batitle{{The Electric Antennas for the STEREO/WAVES Experiment}}.
\bjtitle{Space Sci. Rev.}
\bvolume{136},
\bfpage{529}\,--\,\blpage{547}.
doi:\doiurl{10.1007/s11214-007-9251-x}.
\end{barticle}
\endbibitem

\bibitem[\protect\citeauthoryear{Belheouane}{2014}]{Belheouane14}
\begin{botherref}
\oauthor{\bsnm{Belheouane}, \binits{B.} \bsuffix{Soraya}}:
2014,
{Nanoparticles in the interplanetary medium : spatial observations and theory}.
Theses,
{Universit{\'e} Pierre et Marie Curie - Paris VI}.
\url{https://tel.archives-ouvertes.fr/tel-01086185}.
\end{botherref}
\endbibitem

\bibitem[\protect\citeauthoryear{{Bougeret} \textit{et~al.}}{2008}]{Bougeret08}
\begin{barticle}
\bauthor{\bsnm{{Bougeret}}, \binits{J.L.}},
\bauthor{\bsnm{{Goetz}}, \binits{K.}},
\bauthor{\bsnm{{Kaiser}}, \binits{M.L.}},
\bauthor{\bsnm{{Bale}}, \binits{S.D.}},
\bauthor{\bsnm{{Kellogg}}, \binits{P.J.}},
\bauthor{\bsnm{{Maksimovic}}, \binits{M.}},
\bauthor{\bsnm{{Monge}}, \binits{N.}},
\bauthor{\bsnm{{Monson}}, \binits{S.J.}},
\bauthor{\bsnm{{Astier}}, \binits{P.L.}},
\bauthor{\bsnm{{Davy}}, \binits{S.}},
\bauthor{\bsnm{{Dekkali}}, \binits{M.}},
\bauthor{\bsnm{{Hinze}}, \binits{J.J.}},
\bauthor{\bsnm{{Manning}}, \binits{R.E.}},
\bauthor{\bsnm{{Aguilar-Rodriguez}}, \binits{E.}},
\bauthor{\bsnm{{Bonnin}}, \binits{X.}},
\bauthor{\bsnm{{Briand}}, \binits{C.}},
\bauthor{\bsnm{{Cairns}}, \binits{I.H.}},
\bauthor{\bsnm{{Cattell}}, \binits{C.A.}},
\bauthor{\bsnm{{Cecconi}}, \binits{B.}},
\bauthor{\bsnm{{Eastwood}}, \binits{J.}},
\bauthor{\bsnm{{Ergun}}, \binits{R.E.}},
\bauthor{\bsnm{{Fainberg}}, \binits{J.}},
\bauthor{\bsnm{{Hoang}}, \binits{S.}},
\bauthor{\bsnm{{Huttunen}}, \binits{K.E.J.}},
\bauthor{\bsnm{{Krucker}}, \binits{S.}},
\bauthor{\bsnm{{Lecacheux}}, \binits{A.}},
\bauthor{\bsnm{{MacDowall}}, \binits{R.J.}},
\bauthor{\bsnm{{Macher}}, \binits{W.}},
\bauthor{\bsnm{{Mangeney}}, \binits{A.}},
\bauthor{\bsnm{{Meetre}}, \binits{C.A.}},
\bauthor{\bsnm{{Moussas}}, \binits{X.}},
\bauthor{\bsnm{{Nguyen}}, \binits{Q.N.}},
\bauthor{\bsnm{{Oswald}}, \binits{T.H.}},
\bauthor{\bsnm{{Pulupa}}, \binits{M.}},
\bauthor{\bsnm{{Reiner}}, \binits{M.J.}},
\bauthor{\bsnm{{Robinson}}, \binits{P.A.}},
\bauthor{\bsnm{{Rucker}}, \binits{H.}},
\bauthor{\bsnm{{Salem}}, \binits{C.}},
\bauthor{\bsnm{{Santolik}}, \binits{O.}},
\bauthor{\bsnm{{Silvis}}, \binits{J.M.}},
\bauthor{\bsnm{{Ullrich}}, \binits{R.}},
\bauthor{\bsnm{{Zarka}}, \binits{P.}},
\bauthor{\bsnm{{Zouganelis}}, \binits{I.}}:
\byear{2008},
\batitle{{S/WAVES: The Radio and Plasma Wave Investigation on the STEREO
  Mission}}.
\bjtitle{Space Sci. Rev.}
\bvolume{136},
\bfpage{487}\,--\,\blpage{528}.
doi:\doiurl{10.1007/s11214-007-9298-8}.
\end{barticle}
\endbibitem

\bibitem[\protect\citeauthoryear{{Czechowski} and {Mann}}{2010}]{Czechowski10}
\begin{barticle}
\bauthor{\bsnm{{Czechowski}}, \binits{A.}},
\bauthor{\bsnm{{Mann}}, \binits{I.}}:
\byear{2010},
\batitle{{Formation and Acceleration of Nano Dust in the Inner Heliosphere}}.
\bjtitle{The Astrophysical Journal}
\bvolume{714},
\bfpage{89}\,--\,\blpage{99}.
\end{barticle}
\endbibitem

\bibitem[\protect\citeauthoryear{{Hsu}, {Kr{\"u}ger}, and
  {Postberg}}{2012}]{Hsu12}
\begin{bchapter}
\bauthor{\bsnm{{Hsu}}, \binits{H.-W.}},
\bauthor{\bsnm{{Kr{\"u}ger}}, \binits{H.}},
\bauthor{\bsnm{{Postberg}}, \binits{F.}}:
\byear{2012},
\bctitle{{Dynamics, Composition, and Origin of Jovian and Saturnian Dust-Stream
  Particles}}.
In: \beditor{\bsnm{{Mann}}, \binits{I.}},
\beditor{\bsnm{{Meyer-Vernet}}, \binits{N.}},
\beditor{\bsnm{{Czechowski}}, \binits{A.}} (eds.)
\bbtitle{Nanodust in the Solar System},
\bsertitle{Astrophys. Space Sci. Lib.}
\bseriesno{385},
\bfpage{77}.
doi:\doiurl{10.1007/978-3-642-27543-2\_5}.
\end{bchapter}
\endbibitem

\bibitem[\protect\citeauthoryear{{Hsu} \textit{et~al.}}{2011}]{Hsu11}
\begin{barticle}
\bauthor{\bsnm{{Hsu}}, \binits{H.-W.}},
\bauthor{\bsnm{{Postberg}}, \binits{F.}},
\bauthor{\bsnm{{Kempf}}, \binits{S.}},
\bauthor{\bsnm{{Trieloff}}, \binits{M.}},
\bauthor{\bsnm{{Burton}}, \binits{M.}},
\bauthor{\bsnm{{Roy}}, \binits{M.}},
\bauthor{\bsnm{{Moragas-Klostermeyer}}, \binits{G.}},
\bauthor{\bsnm{{Srama}}, \binits{R.}}:
\byear{2011},
\batitle{{Stream particles as the probe of the dust-plasma-magnetosphere
  interaction at Saturn}}.
\bjtitle{J. Geophys. Res.}
\bvolume{116},
\bfpage{9215}.
doi:\doiurl{10.1029/2011JA016488}.
\end{barticle}
\endbibitem

\bibitem[\protect\citeauthoryear{{Hsu} \textit{et~al.}}{2013}]{Hsu13}
\begin{barticle}
\bauthor{\bsnm{{Hsu}}, \binits{H.-W.}},
\bauthor{\bsnm{{Hansen}}, \binits{K.C.}},
\bauthor{\bsnm{{Hor{\'a}Nyi}}, \binits{M.}},
\bauthor{\bsnm{{Kempf}}, \binits{S.}},
\bauthor{\bsnm{{Mocker}}, \binits{A.}},
\bauthor{\bsnm{{Moragas-Klostermeyer}}, \binits{G.}},
\bauthor{\bsnm{{Postberg}}, \binits{F.}},
\bauthor{\bsnm{{Srama}}, \binits{R.}},
\bauthor{\bsnm{{Zieger}}, \binits{B.}}:
\byear{2013},
\batitle{{Probing IMF using nanodust measurements from inside Saturn's
  magnetosphere}}.
\bjtitle{Geophys. Res. Lett.}
\bvolume{40},
\bfpage{2902}\,--\,\blpage{2906}.
doi:\doiurl{10.1002/grl.50604}.
\end{barticle}
\endbibitem

\bibitem[\protect\citeauthoryear{{Jian} \textit{et~al.}}{2013}]{Jian13}
\begin{bchapter}
\bauthor{\bsnm{{Jian}}, \binits{L.K.}},
\bauthor{\bsnm{{Russell}}, \binits{C.T.}},
\bauthor{\bsnm{{Luhmann}}, \binits{J.G.}},
\bauthor{\bsnm{{Galvin}}, \binits{A.B.}},
\bauthor{\bsnm{{Simunac}}, \binits{K.D.C.}}:
\byear{2013},
\bctitle{{Solar wind observations at STEREO: 2007 - 2011}}.
In: \beditor{\bsnm{{Zank}}, \binits{G.P.}},
\beditor{\bsnm{{Borovsky}}, \binits{J.}},
\beditor{\bsnm{{Bruno}}, \binits{R.}},
\beditor{\bsnm{{Cirtain}}, \binits{J.}},
\beditor{\bsnm{{Cranmer}}, \binits{S.}},
\beditor{\bsnm{{Elliott}}, \binits{H.}},
\beditor{\bsnm{{Giacalone}}, \binits{J.}},
\beditor{\bsnm{{Gonzalez}}, \binits{W.}},
\beditor{\bsnm{{Li}}, \binits{G.}},
\beditor{\bsnm{{Marsch}}, \binits{E.}},
\beditor{\bsnm{{Moebius}}, \binits{E.}},
\beditor{\bsnm{{Pogorelov}}, \binits{N.}},
\beditor{\bsnm{{Spann}}, \binits{J.}},
\beditor{\bsnm{{Verkhoglyadova}}, \binits{O.}} (eds.)
\bbtitle{Am. Inst. Phys. Conf. Ser.}
\bseriesno{CS-1539},
\bfpage{191}\,--\,\blpage{194}.
doi:\doiurl{10.1063/1.4811020}.
\end{bchapter}
\endbibitem

\bibitem[\protect\citeauthoryear{{Jian} \textit{et~al.}}{2006a}]{Jian06a}
\begin{barticle}
\bauthor{\bsnm{{Jian}}, \binits{L.}},
\bauthor{\bsnm{{Russell}}, \binits{C.T.}},
\bauthor{\bsnm{{Luhmann}}, \binits{J.G.}},
\bauthor{\bsnm{{Skoug}}, \binits{R.M.}}:
\byear{2006}a,
\batitle{{Properties of Interplanetary Coronal Mass Ejections at One AU During
  1995 2004}}.
\bjtitle{Solar Phys.}
\bvolume{239},
\bfpage{393}\,--\,\blpage{436}.
doi:\doiurl{10.1007/s11207-006-0133-2}.
\end{barticle}
\endbibitem

\bibitem[\protect\citeauthoryear{{Jian} \textit{et~al.}}{2006b}]{Jian06b}
\begin{barticle}
\bauthor{\bsnm{{Jian}}, \binits{L.}},
\bauthor{\bsnm{{Russell}}, \binits{C.T.}},
\bauthor{\bsnm{{Luhmann}}, \binits{J.G.}},
\bauthor{\bsnm{{Skoug}}, \binits{R.M.}}:
\byear{2006}b,
\batitle{{Properties of Stream Interactions at One AU During 1995 2004}}.
\bjtitle{Solar Phys.}
\bvolume{239},
\bfpage{337}\,--\,\blpage{392}.
doi:\doiurl{10.1007/s11207-006-0132-3}.
\end{barticle}
\endbibitem

\bibitem[\protect\citeauthoryear{{Juh{\'a}sz} and
  {Hor{\'a}nyi}}{2013}]{Juhasz13}
\begin{barticle}
\bauthor{\bsnm{{Juh{\'a}sz}}, \binits{A.}},
\bauthor{\bsnm{{Hor{\'a}nyi}}, \binits{M.}}:
\byear{2013},
\batitle{{Dynamics and distribution of nano-dust particles in the inner solar
  system}}.
\bjtitle{Geophys. Res. Lett.}
\bvolume{40},
\bfpage{2500}\,--\,\blpage{2504}.
doi:\doiurl{10.1002/grl.50535}.
\end{barticle}
\endbibitem

\bibitem[\protect\citeauthoryear{{Le Chat} \textit{et~al.}}{2013}]{LeChat13}
\begin{barticle}
\bauthor{\bsnm{{Le Chat}}, \binits{G.}},
\bauthor{\bsnm{{Zaslavsky}}, \binits{A.}},
\bauthor{\bsnm{{Meyer-Vernet}}, \binits{N.}},
\bauthor{\bsnm{{Issautier}}, \binits{K.}},
\bauthor{\bsnm{{Belheouane}}, \binits{S.}},
\bauthor{\bsnm{{Pantellini}}, \binits{F.}},
\bauthor{\bsnm{{Maksimovic}}, \binits{M.}},
\bauthor{\bsnm{{Zouganelis}}, \binits{I.}},
\bauthor{\bsnm{{Bale}}, \binits{S.D.}},
\bauthor{\bsnm{{Kasper}}, \binits{J.C.}}:
\byear{2013},
\batitle{{Interplanetary Nanodust Detection by the Solar Terrestrial Relations
  Observatory/WAVES Low Frequency Receiver}}.
\bjtitle{Solar Phys.}
\bvolume{286},
\bfpage{549}\,--\,\blpage{559}.
doi:\doiurl{10.1007/s11207-013-0268-x}.
\end{barticle}
\endbibitem

\bibitem[\protect\citeauthoryear{{Lomb}}{1976}]{Lomb76}
\begin{barticle}
\bauthor{\bsnm{{Lomb}}, \binits{N.R.}}:
\byear{1976},
\batitle{{Least-squares frequency analysis of unequally spaced data}}.
\bjtitle{Astrophys. Space Sci.}
\bvolume{39},
\bfpage{447}\,--\,\blpage{462}.
doi:\doiurl{10.1007/BF00648343}.
\end{barticle}
\endbibitem

\bibitem[\protect\citeauthoryear{{Mann}, {Meyer-Vernet}, and
  {Czechowski}}{2014}]{Mann2014}
\begin{barticle}
\bauthor{\bsnm{{Mann}}, \binits{I.}},
\bauthor{\bsnm{{Meyer-Vernet}}, \binits{N.}},
\bauthor{\bsnm{{Czechowski}}, \binits{A.}}:
\byear{2014},
\batitle{{Dust in the planetary system: Dust interactions in space plasmas of
  the solar system}}.
\bjtitle{Phys. Rep.}
\bvolume{536},
\bfpage{1}\,--\,\blpage{39}.
doi:\doiurl{10.1016/j.physrep.2013.11.001}.
\end{barticle}
\endbibitem

\bibitem[\protect\citeauthoryear{{Mann} \textit{et~al.}}{2010}]{Mann2010}
\begin{barticle}
\bauthor{\bsnm{{Mann}}, \binits{I.}},
\bauthor{\bsnm{{Czechowski}}, \binits{A.}},
\bauthor{\bsnm{{Meyer-Vernet}}, \binits{N.}},
\bauthor{\bsnm{{Zaslavsky}}, \binits{A.}},
\bauthor{\bsnm{{Lamy}}, \binits{H.}}:
\byear{2010},
\batitle{{Dust in the interplanetary medium}}.
\bjtitle{Plasma Phys. Controlled Fusion}
\bvolume{52}(\bissue{12}),
\bfpage{124012}.
doi:\doiurl{10.1088/0741-3335/52/12/124012}.
\end{barticle}
\endbibitem

\bibitem[\protect\citeauthoryear{{McBride} and {McDonnell}}{1999}]{McBride99}
\begin{barticle}
\bauthor{\bsnm{{McBride}}, \binits{N.}},
\bauthor{\bsnm{{McDonnell}}, \binits{J.A.M.}}:
\byear{1999},
\batitle{{Meteoroid impacts on spacecraft:sporadics, streams, and the 1999
  Leonids}}.
\bjtitle{Planet. Space Sci.}
\bvolume{47},
\bfpage{1005}\,--\,\blpage{1013}.
doi:\doiurl{10.1016/S0032-0633(99)00023-9}.
\end{barticle}
\endbibitem

\bibitem[\protect\citeauthoryear{{Meyer-Vernet}
  \textit{et~al.}}{2009}]{Meyer09}
\begin{barticle}
\bauthor{\bsnm{{Meyer-Vernet}}, \binits{N.}},
\bauthor{\bsnm{{Maksimovic}}, \binits{M.}},
\bauthor{\bsnm{{Czechowski}}, \binits{A.}},
\bauthor{\bsnm{{Mann}}, \binits{I.}},
\bauthor{\bsnm{{Zouganelis}}, \binits{I.}},
\bauthor{\bsnm{{Goetz}}, \binits{K.}},
\bauthor{\bsnm{{Kaiser}}, \binits{M.L.}},
\bauthor{\bsnm{{St.~Cyr}}, \binits{O.C.}},
\bauthor{\bsnm{{Bougeret}}, \binits{J.-L.}},
\bauthor{\bsnm{{Bale}}, \binits{S.D.}}:
\byear{2009},
\batitle{{Dust Detection by the Wave Instrument on STEREO: Nanoparticles Picked
  up by the Solar Wind?}}
\bjtitle{Solar Phys.}
\bvolume{256},
\bfpage{463}\,--\,\blpage{474}.
doi:\doiurl{10.1007/s11207-009-9349-2}.
\end{barticle}
\endbibitem

\bibitem[\protect\citeauthoryear{{Pantellini}
  \textit{et~al.}}{2012}]{Pantellini12}
\begin{barticle}
\bauthor{\bsnm{{Pantellini}}, \binits{F.}},
\bauthor{\bsnm{{Landi}}, \binits{S.}},
\bauthor{\bsnm{{Zaslavsky}}, \binits{A.}},
\bauthor{\bsnm{{Meyer-Vernet}}, \binits{N.}}:
\byear{2012},
\batitle{{On the unconstrained expansion of a spherical plasma cloud turning
  collisionless: case of a cloud generated by a nanometre dust grain impact on
  an uncharged target in space}}.
\bjtitle{Plasma Physics Controlled Fusion}
\bvolume{54}(\bissue{4}),
\bfpage{045005}.
doi:\doiurl{10.1088/0741-3335/54/4/045005}.
\end{barticle}
\endbibitem

\bibitem[\protect\citeauthoryear{{Pantellini}
  \textit{et~al.}}{2013}]{Pantellini13}
\begin{bchapter}
\bauthor{\bsnm{{Pantellini}}, \binits{F.}},
\bauthor{\bsnm{{Le Chat}}, \binits{G.}},
\bauthor{\bsnm{{Belheouane}}, \binits{S.}},
\bauthor{\bsnm{{Meyer-Vernet}}, \binits{N.}},
\bauthor{\bsnm{{Zaslavsky}}, \binits{A.}}:
\byear{2013},
\bctitle{{On the detection of nano dust using spacecraft based boom antennas}}.
In: \beditor{\bsnm{{Zank}}, \binits{G.P.}},
\beditor{\bsnm{{Borovsky}}, \binits{J.}},
\beditor{\bsnm{{Bruno}}, \binits{R.}},
\beditor{\bsnm{{Cirtain}}, \binits{J.}},
\beditor{\bsnm{{Cranmer}}, \binits{S.}},
\beditor{\bsnm{{Elliott}}, \binits{H.}},
\beditor{\bsnm{{Giacalone}}, \binits{J.}},
\beditor{\bsnm{{Gonzalez}}, \binits{W.}},
\beditor{\bsnm{{Li}}, \binits{G.}},
\beditor{\bsnm{{Marsch}}, \binits{E.}},
\beditor{\bsnm{{Moebius}}, \binits{E.}},
\beditor{\bsnm{{Pogorelov}}, \binits{N.}},
\beditor{\bsnm{{Spann}}, \binits{J.}},
\beditor{\bsnm{{Verkhoglyadova}}, \binits{O.}} (eds.)
\bbtitle{Am. Inst. Phys. Conf. Ser.}
\bseriesno{CS-1539},
\bfpage{414}\,--\,\blpage{417}.
doi:\doiurl{10.1063/1.4811073}.
\end{bchapter}
\endbibitem

\bibitem[\protect\citeauthoryear{{Scargle}}{1982}]{Scargle82}
\begin{barticle}
\bauthor{\bsnm{{Scargle}}, \binits{J.D.}}:
\byear{1982},
\batitle{{Studies in astronomical time series analysis. II - Statistical
  aspects of spectral analysis of unevenly spaced data}}.
\bjtitle{The Astrophys. J.}
\bvolume{263},
\bfpage{835}\,--\,\blpage{853}.
doi:\doiurl{10.1086/160554}.
\end{barticle}
\endbibitem

\bibitem[\protect\citeauthoryear{{Schippers}
  \textit{et~al.}}{2014}]{Schippers14}
\begin{barticle}
\bauthor{\bsnm{{Schippers}}, \binits{P.}},
\bauthor{\bsnm{{Meyer-Vernet}}, \binits{N.}},
\bauthor{\bsnm{{Lecacheux}}, \binits{A.}},
\bauthor{\bsnm{{Kurth}}, \binits{W.S.}},
\bauthor{\bsnm{{Mitchell}}, \binits{D.G.}},
\bauthor{\bsnm{{Andr{\'e}}}, \binits{N.}}:
\byear{2014},
\batitle{{Nanodust detection near 1 AU from spectral analysis of Cassini/RPWS
  radio data}}.
\bjtitle{Geophys. Res. Lett.}
\bvolume{41},
\bfpage{5282}\,--\,\blpage{5388}.
doi:\doiurl{10.1002/2014GL060566}.
\end{barticle}
\endbibitem

\bibitem[\protect\citeauthoryear{{Townsend}}{2010}]{Townsend10}
\begin{barticle}
\bauthor{\bsnm{{Townsend}}, \binits{R.H.D.}}:
\byear{2010},
\batitle{{Fast Calculation of the Lomb-Scargle Periodogram Using Graphics
  Processing Units}}.
\bjtitle{The Astrophys. J. Supp.}
\bvolume{191},
\bfpage{247}\,--\,\blpage{253}.
doi:\doiurl{10.1088/0067-0049/191/2/247}.
\end{barticle}
\endbibitem

\bibitem[\protect\citeauthoryear{{Zaslavsky}
  \textit{et~al.}}{2012}]{Zaslavsky12}
\begin{barticle}
\bauthor{\bsnm{{Zaslavsky}}, \binits{A.}},
\bauthor{\bsnm{{Meyer-Vernet}}, \binits{N.}},
\bauthor{\bsnm{{Mann}}, \binits{I.}},
\bauthor{\bsnm{{Czechowski}}, \binits{A.}},
\bauthor{\bsnm{{Issautier}}, \binits{K.}},
\bauthor{\bsnm{{Le Chat}}, \binits{G.}},
\bauthor{\bsnm{{Pantellini}}, \binits{F.}},
\bauthor{\bsnm{{Goetz}}, \binits{K.}},
\bauthor{\bsnm{{Maksimovic}}, \binits{M.}},
\bauthor{\bsnm{{Bale}}, \binits{S.D.}},
\bauthor{\bsnm{{Kasper}}, \binits{J.C.}}:
\byear{2012},
\batitle{{Interplanetary dust detection by radio antennas: mass calibration and
  fluxes measured by STEREO/WAVES}}.
\bjtitle{J. Geophys. Res.}
\bvolume{117},
\bfpage{5102}.
doi:\doiurl{10.1029/2011JA017480}.
\end{barticle}
\endbibitem

\end{thebibliography}
%
%
%
%

\end{article} 
\end{document}